\documentclass[twocolumn, pra, floatfix, superscriptaddress, aps, showpacs]{revtex4-1}

\usepackage{graphicx}
\usepackage{tikz}
\usepackage{helvet}

\begin{document}

\title{Collective strong coupling of cold atoms to an all-fiber ring cavity}

\author{S.~K.~Ruddell}\email{sam.ruddell@auckland.ac.nz}\affiliation{The Dodd-Walls Centre for Photonic and Quantum Technologies}\affiliation{Department of Physics, University of Auckland, Private Bag 92019, Auckland, New Zealand}

\author{K.~E.~Webb}\affiliation{The Dodd-Walls Centre for Photonic and Quantum Technologies}\affiliation{Department of Physics, University of Auckland, Private Bag 92019, Auckland, New Zealand}

\author{I.~Herrera}\affiliation{The Dodd-Walls Centre for Photonic and Quantum Technologies}\affiliation{Department of Physics, University of Auckland, Private Bag 92019, Auckland, New Zealand}

\author{A.~S.~Parkins}\affiliation{The Dodd-Walls Centre for Photonic and Quantum Technologies}\affiliation{Department of Physics, University of Auckland, Private Bag 92019, Auckland, New Zealand}

\author{M.~D.~Hoogerland}\affiliation{The Dodd-Walls Centre for Photonic and Quantum Technologies}\affiliation{Department of Physics, University of Auckland, Private Bag 92019, Auckland, New Zealand}

\begin{abstract}
We experimentally demonstrate a ring geometry all-fiber cavity system for cavity quantum electrodynamics with an ensemble of cold atoms. The fiber cavity contains a nanofiber section which mediates atom-light interactions through an evanescent field. We observe well-resolved, vacuum Rabi splitting of the cavity transmission spectrum in the weak driving limit due to a collective enhancement of the coupling rate by the ensemble of atoms within the evanescent field, and we present a simple theoretical model to describe this. In addition, we demonstrate a method to control and stabilize the resonant frequency of the cavity by utilizing the thermal properties of the nanofiber.
\end{abstract}

\maketitle

\noindent 
In the context of scalable quantum computing and quantum communication, one can envision a network of quantum systems, or nodes, 
linked together through appropriate quantum channels to form a quantum network \cite{Chou2007, Kimble2008, Northup2014}. These networks have a wide variety of applications, from quantum computation \cite{Ladd2010} to simulating many-body quantum systems \cite{Georgescu2014}.
The use of photons for quantum communication is well established \cite{Gisin2007}, and hence quantum nodes require an efficient light-matter interface. Cavity quantum electrodynamics (QED) has been at the forefront of implementing strong, coherent atom-light interactions \cite{Miller2005,Wilk2007,Haroche2013}, and the ability of cavity QED systems to control and manipulate the quantum states of light and matter make them perfect candidates for quantum nodes \cite{Reiserer2015}. For example, free-space Fabry-P\'erot cavities have been used to implement an elementary quantum network consisting of two nodes \cite{Ritter2012}. 

To overcome the complexity and poor scalability of free-space cavities, fiber-based alternatives are required in order to realize a large scale quantum network. When implementing an all-fiber quantum node, a useful component for achieving strong coupling between fiber guided light and atoms is an optical nanofiber. Such a nanofiber offers tight transversal-mode confinement and large evanescent fields, and therefore allows for efficient atom-light coupling \cite{Nayak2007,Vetsch2010}, as well as nonlinear atom-light interactions at very low optical powers \cite{Spillane2008}. Recently, these nanofibers have been utilized to demonstrate a memory for light \cite{Sayrin2015,Gouraud2015}. 

The atom-light interaction can be further enhanced by the use of a fiber cavity. One such configuration utilizes a nanofiber section sandwiched between two fiber-Bragg gratings to form an all-fiber Fabry-P\'erot cavity \cite{Wuttke2012}. Strong optical coupling with a single cesium atom has been demonstrated in this system, with a moderate cavity finesse \mbox{($\mathcal{F}<40$)}, due to high quality trapping of the atom near to the nanofiber surface  \cite{Kato2015}. 

An alternative ring geometry has been recently studied using a low finesse fiber cavity with a hot atomic vapor of rubidium, where it was seen that the cavity transmission varied as the probe frequency was scanned across the Doppler-broadened atomic resonance \cite{Jones2016}. Furthermore, in a recent proposal, a ring cavity geometry has been suggested as a potential platform for studying multimode strong optical coupling and chiral cavity QED \cite{Schneeweiss2017}. 

Here, we report on the first demonstration of the use of a fiber ring cavity containing a nanofiber section to observe collectively enhanced strong coupling of an ensemble of atoms to a mode of the cavity. We achieve a moderate finesse of $\mathcal{F}\approx35$ by splicing the untapered ends of a nanofiber section together with a fiber-optical coupler. By utilizing the thermal properties of the nanofiber, we are able to stabilize the cavity resonance to the atomic transition. The guided mode of the nanofiber is allowed to interact with a cold cloud of cesium atoms, and we observe a splitting of the cavity resonance due to a collective enhancement by the ensemble of atoms in the weak driving limit. We present a simple theoretical model to describe this splitting, as well as saturation of the atoms at higher driving strengths. 

\begin{figure}[t]
\centering
\begin{tikzpicture}[>=latex, scale=1.2]

\draw [line width=3pt, gray] (0,0.9) circle (1);

\shade[shading=radial, inner color=red!70!orange, outer color=white] (0,0.9) circle (0.3);

\draw[thick] (-0.5,0.9) -- (0.5,0.9);
\node[above] at (0,1.15){\textsf{cesium}};
\node[above] at (0.15,0.2){\textsf{nanofiber}};
\draw[->](0.45,0.5) -- (0.4,0.85);

\draw[ultra thick,domain=0:180] plot ({0.6*sin(\x) + 1}, {0.7*cos(\x)+0.2});
\draw[ultra thick,domain=180:360] plot ({0.6*sin(\x) - 1}, {0.7*cos(\x)+0.2});
\draw[ultra thick,->](1.6,0.3) -- (1.6,0.38) ;	
\draw[ultra thick] (-1,0.9) -- (-0.5,0.9) ;
\draw[ultra thick] (0.5,0.9)--(1,0.9) ;
\draw[ultra thick, rounded corners=0.1] (-1,-0.5) -- (-0.5,-0.5) -- (-0.5,-0.525) -- (0.5,-0.525) -- (0.5,-0.5) -- (1,-0.5);
\draw[ultra thick, rounded corners=0.1] (-1.8,-0.55) -- (-0.5,-0.55) -- (-0.5,-0.525) -- (0.5,-0.525) -- (0.5,-0.55) -- (1.8,-0.55);

\filldraw[fill=gray, opacity=0.95, rounded corners=1](-0.55,-0.45) rectangle (0.55,-0.6);
\node[below=4] at (0,-0.55){\textsf{fiber coupler}};

\draw[very thick, red!70!orange, rounded corners=0.3] (-3.2, 0.85) -- (-3.2,-0.564) -- (-2.348,-0.564) -- (-2.3,-0.55) -- (-1.8,-0.55);
\draw[thick, red!70!orange, ->] (-3.2, 0.85) -- (-3.2,0.55);
\draw[thick, blue, ->] (-2.3, 0.85) -- (-2.3,0.55);	
\draw[thick, blue] (-1.8,-0.55) -- (-2.3,-0.55);
\draw[thick, blue] (-2.3,-0.55)-- (-2.3,0.85);

\draw[very thick, red!70!orange] (1.8,-0.55) -- (2.3,-0.55);
\draw[very thick, red!70!orange] (2.3,-0.55) -- (2.348,-0.564) -- (3.2,-0.564);
\draw[very thick, red!70!orange,->] (2.348,-0.564) -- (2.78,-0.564);
\draw[thick, blue] (1.8,-0.55) -- (2.3,-0.55);
\draw[thick, blue] (2.3,-0.55) -- (2.3,0.45);
\draw[thick, blue, ->] (2.3,-0.55) -- (2.3,0);

\draw[thick](-2.55,0.85) rectangle (-2.05,1.659);
\node[below](piezo) at (-2.3,1.659){};
\node[shift={(0,-0.2)}] at (piezo){\textsf{780}};
\node[shift={(0,-0.5)}] at (piezo){\textsf{nm}};

\draw[thick](-3.45,0.85) rectangle (-2.95,1.659);
\node[below](piezo2) at (-3.2,1.659){};
\node[shift={(0,-0.2)}] at (piezo2){\textsf{852}};
\node[shift={(0,-0.5)}] at (piezo2){\textsf{nm}};

\begin{scope}[shift={(-3.2,-0.564)},rotate={225}]
	\filldraw[fill=gray, opacity=0.7](0.05,0.25) -- (0.05,-0.25) -- (0,-0.25) -- (0,0.25) -- cycle;
\end{scope}

\draw[very thick](2.1,0.45) -- (2.5,0.45); 

\begin{scope}[shift={(3.2,-0.564)},scale={0.01}]
	\draw[thick](0,15)--(0,-15)--(2.2,-15)--(2.2,15)--cycle;
	\draw[thick, rounded corners=1](2.2,-19.5)--(27.6,-19.5)--(27.6,44)--(2.2,44)--cycle;
	\node[below] (apd) at (13.8,44){};
	\node[shift={(0,-0.9)}] at (apd) {\textsf{APD}};
\end{scope}

\draw[very thick, dashed, rounded corners=10, ->] (apd) -- ++(0,2.4) -- ++(-2.92,0) -- ++(-2.72,0) -- (piezo);
\node[shift={(0.9,0.5)}] at (piezo){\textsf{feedback}};	

\begin{scope}[shift={(-2.3,-0.55)},rotate={225}]
	\filldraw[fill=gray, opacity=0.7](0.05,0.25) -- (0.05,-0.25) -- (0,-0.25) -- (0,0.25) -- cycle;
	\node[below=8] at (0,0){\textsf{DM}};
\end{scope}

\begin{scope}[shift={(2.3,-0.55)},rotate={-45}]
	\filldraw[fill=gray, opacity=0.7](0.05,0.25) -- (0.05,-0.25) -- (0,-0.25) -- (0,0.25) -- cycle;
	\node[below=8] at (0,0){\textsf{DM}};
\end{scope}

\begin{scope}[shift={(2.85,-0.55)}]
	\filldraw[fill=gray, opacity=0.7](0.05,0.25) -- (0.05,-0.25) -- (0,-0.25) -- (0,0.25) -- cycle;
	\node[above=8] at (0,0){\textsf{filter}};
\end{scope}

\begin{scope}[shift={(-3.2,-0.1)},rotate={90}]
	\filldraw[fill=gray, opacity=0.7](0.05,0.25) -- (0.05,-0.25) -- (0,-0.25) -- (0,0.25) -- cycle;
	\node[shift={(0.6,0.025)}]{$\lambda/2$};
\end{scope}
\begin{scope}[shift={(-3.2,0.2)},rotate={90}]
	\filldraw[fill=gray, opacity=0.7](0.05,0.25) -- (0.05,-0.25) -- (0,-0.25) -- (0,0.25) -- cycle;
	\node[shift={(0.6,0.025)}]{$\lambda/4$};
\end{scope}

\begin{scope}[shift={(0.85,1.4)}]
	\node[right] at (0,0.35){\textsf{vacuum}};
	\node[right] at (0,0.13){\textsf{chamber}};
\end{scope}

\end{tikzpicture}
\caption{Schematic of the experimental setup. We insert two lasers into the cavity system, an 852~nm probe laser and a 780~nm heating laser. The lasers are combined and split using dichroic mirrors (DM), and the probe laser output is measured using an avalanche photodiode (APD). This signal is fed back to the 780~nm laser frequency in order to control the position of the cavity resonance.}
\label{fig:experiment}
\end{figure}
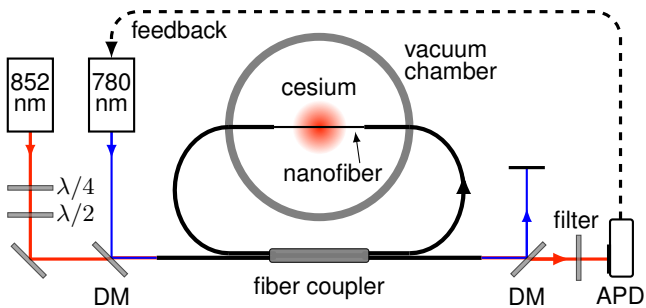

The schematic of the experimental setup is shown in Fig.~\ref{fig:experiment}. Our cavity utilizes a section of fiber that has been tapered down to a nanofiber, having a waist diameter of $\sim$400~nm and a waist length of $\sim$2~mm. The nanofiber is fabricated using the flame-brush method \cite{Bilodeau1988,Birks1992}, with a taper profile optimized to minimize its length while maintaining a high transmission, as described in \cite{Nagai2014}. Using this method we are able to achieve nanofiber transmissions exceeding 99\%. The nanofiber is then placed inside a vacuum chamber, and the untapered fiber ends are fed through to the outside via fiber feedthroughs \cite{Abraham1998}. The fiber ring cavity is then formed by splicing together the ends of the fiber with a 95:5 fused fiber coupler. Measurements of the transmission through a single splice show typical losses of approximately 1\%. 

\begin{figure}[t]
\centering
\includegraphics[width=83mm]{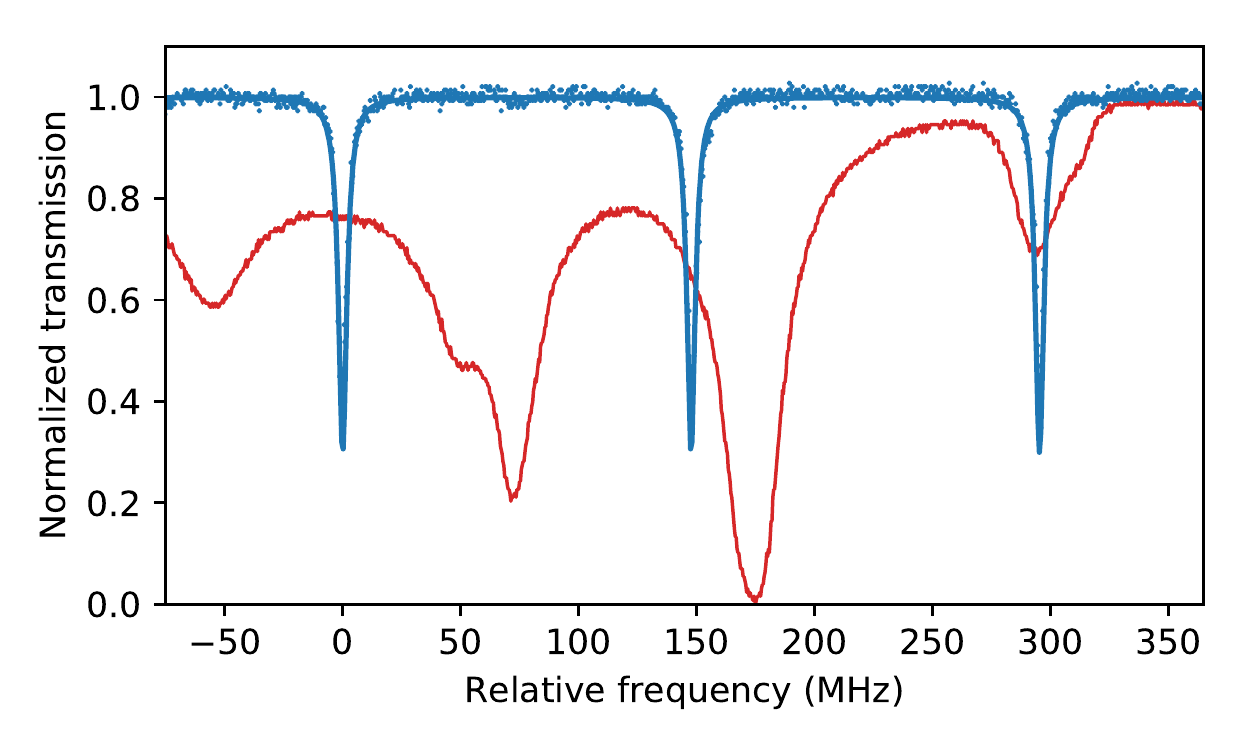}
\caption{Characterization of the empty cavity. We fit the cavity output using a simple theoretical model \cite{Heebner2004}, shown in blue, and obtain a finesse $\mathcal{F}\approx35$ and a free spectral range of $148\pm1$~MHz. An absorption spectroscopy signal from a reference cesium cell is also shown in red, and provides the calibration of the frequency scale.}
\label{fig:cavRes}
\end{figure}

We characterize the cavity by injecting a weak probe laser at 852~nm into the cavity through the fiber coupler, and monitor the cavity resonances as the probe laser frequency is scanned over multiple free-spectral ranges (FSRs). The cavity transmission for a laser scan over approximately three FSRs is shown in Fig.~\ref{fig:cavRes}. We fit the cavity transmission following \cite{Heebner2004}, from which we obtain a finesse of $\mathcal{F}\approx35$, and a FSR of 148$\pm$1~MHz, corresponding to a cavity length of 1.4~m. The total field decay rate of the cavity is given by $\kappa = \kappa_{\mathrm{i}} + \kappa_{\mathrm{ex}}$, where $\kappa_{\mathrm{i}}$ corresponds to the intrinsic (or internal) cavity loss rate, and $\kappa_{\mathrm{ex}}$ to the field decay rate through the fiber coupler. We find that $\kappa_{\mathrm{i}}/(2\pi)=1.7$~MHz and $\kappa_{\mathrm{ex}}/(2\pi)=0.47$~MHz, corresponding to a higher intrinsic cavity loss than expected from our measurements of splice loss and nanofiber transmission following fabrication. This is likely due to degradation of the nanofiber by contaminants adsorbed onto the nanofiber surface, and places our system in the undercoupled regime ($\kappa_{\mathrm{i}}>\kappa_{\mathrm{ex}}$).

By utilizing the thermal response of the nanofiber \cite{Wuttke2013}, we are able to stabilize the resonant frequency of the cavity.
As a laser field propagates through the nanofiber, a small fraction of the optical power is absorbed. In vacuum, this power absorption can cause significant heating of the nanofiber with sufficient laser intensity. This in turn induces a change in refractive index of the nanofiber as well as a thermal expansion, increasing the optical path length, and shifting the cavity resonance towards lower frequencies. 
By scanning a laser from high to low frequency into a cavity resonance, thereby increasing the intracavity power, it is possible to attain a thermal equilibrium where small perturbations of the laser frequency will be compensated for by a shift in the cavity resonance. This thermal self-stability effect is analogous to that observed in microresonators \cite{Carmon2004}. 

For thermal control of the cavity, we couple $\sim$2~mW of 780~nm laser power, far blue-detuned from the atomic resonance, into the cavity system. We monitor the cavity resonances using 140~nW of 852~nm probe laser power, red-detuned from the atomic resonance. At this power and detuning, the empty cavity lineshape is unchanged in the presence of atoms. This signal is used as feedback for the 780~nm laser frequency, as shown in Fig. \ref{fig:experiment}, and allows us to actively stabilize the cavity to the atomic resonance. In addition, the heating of the nanofiber by the 780~nm laser assists in preventing the contamination and subsequent degradation of the nanofiber transmission by the cesium atoms.

To experimentally investigate atom-light interactions in the fiber ring resonator, we first cool and trap cesium atoms in a standard configuration magneto-optical trap (MOT) positioned to overlap the nanofiber section. For cooling, we operate a laser detuned from the $D_2$-line $F=4\rightarrow F^\prime=5$ transition by $-2.6\Gamma$ with an intensity of $6I_\mathrm{s}$, where $\Gamma$ is the natural linewidth and $I_\mathrm{s}$ is the saturation intensity of the transition. A second repumping laser is tuned to the $F=3\rightarrow F^\prime=4$ transition and overlapped with the cooling beams. For a standard MOT density of approximately $10^{10}\,{\rm cm}^{-3}$, this corresponds to roughly $100$ atoms interacting with the cavity mode through the evanescent field. At this stage, the cavity is stabilized and locked to the atomic resonance using the thermal self-stabilization effect described previously. 

An experimental sequence begins by turning off the probe and MOT cooling lasers. As the active stabilization of the cavity relies on feedback from the probe laser, we also pause the cavity stabilization at this point. The repump laser is turned off 500~$\mu$s later to ensure that all atoms are pumped into the $F=4$ ground state. After another 500~$\mu$s, we probe the atoms through the nanofiber with a 1~ms pulse from the probe laser at the desired input power and frequency, before reinstating the MOT lasers and the active cavity stabilization. This sequence is repeated every 100~ms, and the trace is averaged 128 times on an oscilloscope for each data point.

\begin{figure}[t]
\centering
\includegraphics[width=83mm]{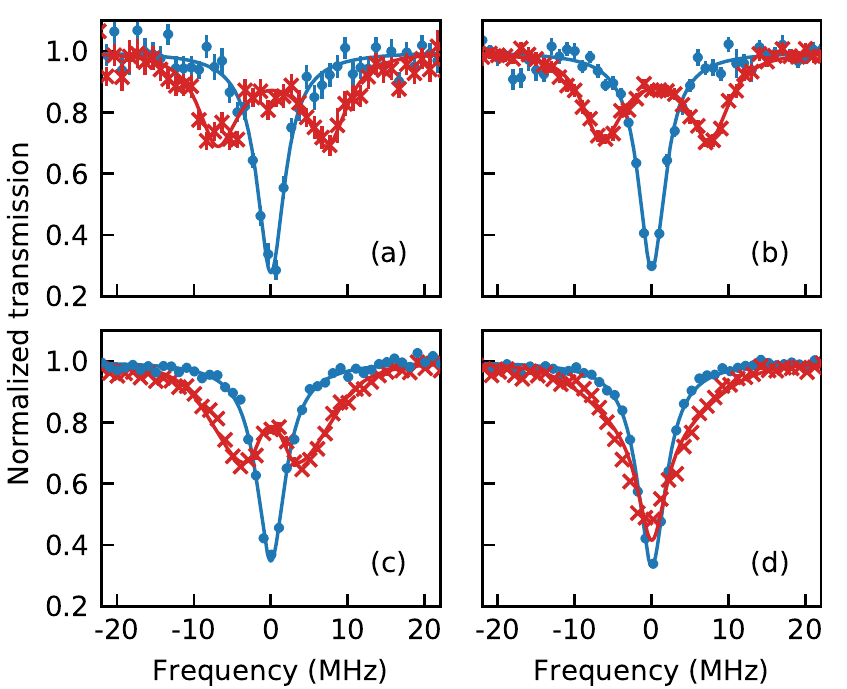}
\caption{Normalized transmission of the cavity as the probe laser frequency is scanned across the atomic resonance for varying input powers $P_\mathrm{in}$. Blue circles show data for a cavity in the absence of atoms with a Lorentzian fit. The red crosses correspond to an ensemble of atoms interacting with the cavity mode, with the theoretical fit given by Eq.~(\ref{eq:Tsat}). Parameters for the fit are $C=1.5$, $\gamma_\perp/(2\pi)=4$~MHz (a) $P_\mathrm{in}=30$~pW, (b) $P_\mathrm{in}=60$~pW, (c) $P_\mathrm{in}=750$~pW, (d) $P_\mathrm{in}=2.3$~nW. Error bars correspond to the standard deviation of each point measured over the 1~ms probe time.}
\label{fig:freqScan}
\end{figure}

To establish the steady-state frequency response of the system, we scan the probe laser frequency with the cavity resonance aligned to the atomic resonance. For low input powers ($P_\mathrm{in}\lesssim$1~nW), we observe a splitting of the cavity resonance, as shown in Figs. \ref{fig:freqScan}(a)--(c). This splitting is a many atom version of the vacuum-Rabi splitting observed with a single atom in \cite{Kato2015}. In our experiment the average single atom-cavity coupling rate is not high enough to cause the splitting alone. Instead, we attribute the observed splitting to a collective enhancement by an ensemble of atoms interacting with the cavity mode, as described below. 

The normalized transmission, derived via the input-output formalism of optical cavities (see, e.g., \cite{Parkins2014}), can be written in the form
\begin{equation}\label{eq:Tsat}
T = \left| 1-\frac{2i}{y}\frac{\kappa_\mathrm{ex}}{\kappa}X \right|^2,	
\end{equation}
where $X$, proportional to the intracavity field amplitude, is determined from the semiclassical, steady state equation 
\begin{equation}\label{eq:T}
y = iX \left( 1 + i\Delta_c + \frac{4C(1-i\Delta_a)}{1+\Delta_a^2+2|X|^2} \right) .
\end{equation}
Here, $y$ is a dimensionless driving field strength, with $|y|^2$ proportional to the ratio of the intracavity photon number with no atoms and resonant driving, to the saturation photon number, $n_{\rm sat}=\gamma_\perp\gamma_\parallel/4g_{\rm eff}^2$, where $g_{\rm eff}$ is an effective (or average) single-atom coupling strength to the cavity mode. Note that $y$ can be related to the input power, $P_{\rm in}$, by
\begin{equation}
|y|^2 = \frac{P_{\rm in}}{2\kappa n_{\rm sat}} \frac{2\kappa_{\rm ex}}{\kappa} \frac{\lambda_p}{2\pi\hbar c} ,
\end{equation}
where $\lambda_p=852$~nm.
The parameters $\Delta_c = \Delta_{\mathrm{cavity}}/\kappa$ and $\Delta_a = \Delta_{\mathrm{atom}}/\gamma_\perp$ are the normalized detunings of the probe laser frequency from the cavity and atomic resonances, respectively, while $\gamma_\perp =\gamma_\parallel /2+\gamma_d$ is the transverse atomic decay rate, with $\gamma_\parallel /(2\pi )=5.2$~MHz and $\gamma_d$ the dephasing rate (due, e.g., to laser frequency noise and the position-dependent light shift of the atoms induced by the 780~nm laser). Finally, the dimensionless measure of the atom-cavity coupling strength is given by the cooperativity,
\begin{equation}
C = \sum_j \frac{g_j^2(\mathbf{r}_j)}{2\kappa\gamma_\perp} ,
\end{equation}
where $g_j(\mathbf{r}_j)$ is the dipole coupling strength of atom $j$ to the cavity field.
Strong coupling between the cavity and the ensemble of atoms occurs when $C\gtrsim 1$, and in the weak driving limit ($|y|\lesssim 1$) the transmission spectrum,
\begin{equation}\label{eq:Tweak}
T_{\mathrm{weak}} \simeq \left| 1 - \frac{2\kappa_{\rm ex}/\kappa}{1+i\Delta_c + \frac{4C(1-i\Delta_a)}{1+\Delta_a^2}} \right|^2,
\end{equation}
exhibits a (peak-to-peak) splitting on the order of $4\sqrt{\kappa\gamma_\perp C}$.


\begin{figure}[t]
\centering
\includegraphics[width=83mm]{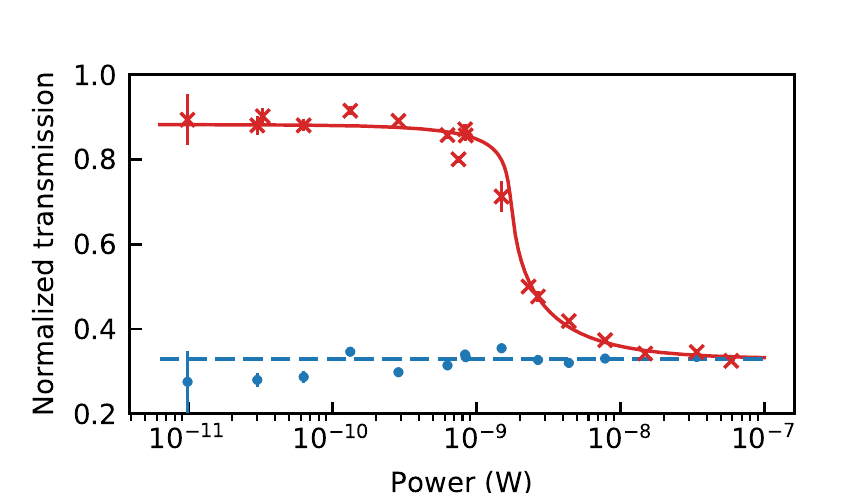}
\caption{ Normalized transmission as a function of input power for the system on resonance, such that $\Delta_a = \Delta_c = 0$. Red crosses correspond to atoms interacting with the cavity mode, and blue dots to an empty cavity. The theoretical curves are calculated using Eq.~(\ref{eq:Tsat}) with $C=1.5, n_{\mathrm{sat}}=12.7, \kappa_{\mathrm{ex}}/(2\pi)=0.47~\mathrm{MHz}$, and $\kappa_\mathrm{i}/(2\pi) = 1.7~\mathrm{MHz}$. Error bars correspond to the standard deviation of each point measured over the 1~ms probe time.}
\label{fig:powerScan}
\end{figure}

We fit the spectra in Figs. \ref{fig:freqScan}(a)--(d) using Eq.~(\ref{eq:Tsat}), where we have set $\Delta_{\rm atom} = \Delta_{\rm cavity}$, corresponding to having the cavity resonance aligned with the atomic resonance. The parameters used for the fits are \mbox{$C=1.5$}, and $\gamma_\perp/(2\pi)=4$~MHz. 
We observe a clear splitting of the cavity resonance in the weak driving limit. As the input power $P_\mathrm{in}$ is increased, the splitting of the cavity resonance is reduced, before eventually disappearing. The cavity lineshape then approaches that of the empty cavity. 

To further investigate the power dependence of the system, we perform the same experiment for various powers $P_\mathrm{in}$ on resonance, such that $\Delta_a = \Delta_c = 0$. The response of the system is shown in Fig.~\ref{fig:powerScan}. We observe a distinct nonlinear behavior in the response of the cavity in the presence of atoms, contrasting the linear response of the empty cavity. We again use Eq.~(\ref{eq:Tsat}) to fit the data, with $\{C,\kappa_{\rm ex}, \kappa_{\rm i}, \gamma_\perp\} $ as quoted above,
from which we obtain a saturation photon number of $n_\mathrm{sat}=13\pm1$, which gives the approximate intracavity photon number at which nonlinear effects set in.

We note that the theoretical model does not fully take into account the spatial profile of the cavity mode in the evanescent region, so this value for $n_\mathrm{sat}$ is most likely an over-estimate of the true value. Nevertheless, using $n_\mathrm{sat}=13$ gives $g_{\rm eff}/(2\pi)\simeq 0.6~{\rm MHz}$, and, writing $C=N_{\rm eff}g_{\rm eff}^2/(2\kappa\gamma_\perp)$, yields an effective atom number $N_{\rm eff}\simeq 64$, which appears consistent with the experimental configuration. 

In conclusion, we have constructed a fiber ring cavity containing a nanofiber section to allow for interactions between a cavity mode and a cold cloud of cesium atoms. We achieve a sufficiently high finesse to observe a splitting of the cavity mode resonance due to a collective enhancement of the coupling strength by an ensemble of atoms. We take advantage of the thermal properties of the nanofiber in vacuum to actively stabilize the resonant frequency of the cavity. Our system has the advantage of being simpler to implement than alternative systems utilizing free-space optics, making it an excellent candidate for creating large-scale quantum networks. 

Future experiments will involve trapping atoms near to the nanofiber surface in a dipole trap in order to further enhance the cooperativity \cite{Vetsch2010,Kato2015} and potentially achieve strong coupling with a single atom, which would enable the implementation of elementary quantum logic gates \cite{Reiserer2014,Hacker2016}. The all-fiber nature of the set up should also allow us to explore techniques to increase the stability and fidelity of the system, such as passive feedback via additional fiber loops \cite{Hein2015,Grimsmo2015,Grimsmo2014}, and to create a rudimentary quantum network by the addition of a second node. This in turn opens up the possibility of experiments that explore, for example, superradiant emission of distantly-separated atoms \cite{LeKien2005,Zeeb2015}, quantum state transfer \cite{Cirac1997,Duan2001}, and the preparation of quantum entangled states of separated atoms \cite{Clark2003, Parkins2006, Stannigel2012}.

We thank Takao Aoki for helpful discussions. IH would like to acknowledge funding from the Marsden Fund.

\end{document}